\documentclass[pdflatex,sn-mathphys-num]{sn-jnl}

\usepackage{tikz} 
\usepackage{graphicx}%
\usepackage{multirow}%
\usepackage{amsmath,amssymb,amsfonts}%
\usepackage{amsthm}%
\usepackage{mathrsfs}%
\usepackage[title]{appendix}%
\usepackage{xcolor}%
\usepackage{textcomp}%
\usepackage{manyfoot}%
\usepackage{booktabs}%
\usepackage{algorithm}%
\usepackage{algorithmicx}%
\usepackage{algpseudocode}%
\usepackage{listings}%
\usepackage{float}
\usepackage{siunitx}
\usepackage{bm}
\usepackage{caption}
\usepackage[dvipsnames]{xcolor}

\theoremstyle{thmstyleone}%
\theoremstyle{thmstyletwo}%

\theoremstyle{thmstylethree}%
\newtheorem{definition}{Definition}%

\begin{document}

\title[Article Title]{
Addressing Ill-conditioning in Density Functional Theory for Reliable Machine Learning
}


\author[1]{\fnm{L.} \sur{Arnstein}}

\author[1,2]{\fnm{J.} \sur{Wetherell}}

\author[1]{\fnm{R.} \sur{Lawrence}}

\author[1]{\fnm{P. J.} \sur{Hasnip}}\email{phil.hasnip@york.ac.uk}

\author*[1,2]{\fnm{M. J. P.} \sur{Hodgson}}\email{matt.hodgson@york.ac.uk}

\affil[1]{\orgdiv{School of Physics, Engineering and Technology}, \orgname{The University of York}, \orgaddress{\street{Heslington}, \city{York}, \postcode{YO10 5DD}, \country{UK}}}

\affil[2]{\orgdiv{The European Theoretical Spectroscopy Facility}}


\abstract{In principle, machine learning (ML) can be used to obtain any electronic property of a many-body system from its electron density within density functional theory. However, some physical quantities are highly sensitive to small variations in the density. This `ill-conditioning' limits the accuracy with which these quantities can be learned as density functionals from a fixed amount of data. We identify sources of ill-conditioning present in density functionals that belong to two ubiquitous classes: 1) Physical quantities that are globally gauge-dependent, meaning they change value if a constant shift is applied to the external potential -- for example, the total energy; 2) Functionals of the $N$-electron density that have an implicit dependence on the $(N+1)$-electron density, such as the fundamental gap. We demonstrate that widely used ML models exhibit orders-of-magnitude greater error when applied to these ill-conditioned density functionals compared to other functionals that fall into neither class, even when the global gauge is fixed to prevent constant shifts. Owing to an absence of ill-conditioning in potential functionals, we find that providing the external potential as input to the ML model leads to significantly improved predictions of quantities in these two classes.
}





\maketitle
\clearpage
    
\section{Introduction}\label{sec1}

Due to its balance of accuracy and computational efficiency, density functional theory (DFT) has become the most popular method for electronic structure calculations \cite{becke2014perspective}. DFT replaces the usually incalculable many-body wavefunction with the ground-state electron density -- from which any many-body electronic property can, in principle, be calculated exactly \cite{Hohenberg1964}. In practice, while approximate functional forms have been found for some physical quantities, many others remain completely unknown \cite{Kolb2017}. 

Recently, machine learning (ML) has emerged as a promising approach for developing density functional approximations (DFAs) \cite{Akashi2025}. ML models learn the relationship between the electron density, $n(\bm{r})$, and a physical quantity, $Q$, via a collection of examples known as training data. A successfully trained model should then be able to generalise to unseen data, accurately predicting the values of $Q[n]$ corresponding to densities that were not used for training. These predictions can be made significantly faster than the original method used to generate the training data \cite{Chandrasekaran2019,Ryczko2019}, even when non-local descriptors are used.

Currently, practical DFT calculations employ the Kohn-Sham (KS) approach \cite{KS1965}, which uses an auxiliary system of non-interacting electrons with, in principle, the same electron density as the real, many-body system. This approach relies on approximate forms of the exchange-correlation energy, \(E_\mathrm{xc}[n]\), as the exact functional form is unknown. Although many existing approximations perform well for specific system classes, the search for a universally applicable \(E_\mathrm{xc}[n]\) functional is ongoing \cite{Wu2023}.

The originally intended role of the KS system was to provide the density, which could then be used to calculate any electronic property via the appropriate DFA. However, owing to a lack of reliable explicit DFAs for many physical quantities, the electronic properties of the KS system are often used instead, despite known limitations. For example, the KS band (fundamental) gap is often used in place of the real, many-body band gap, even when they differ significantly \cite{perdew1985density,Godby1988}. In an effort to overcome these limitations, there is growing interest in constructing ML-DFAs for a broader range of electronic properties, such as the band gap \cite{Kolb2017,Moreno2021,Gedeon_2021,Pilania2013,Chen2025} and the electron affinity \cite{Pilania2013}. 
In addition, provided suitably accurate models can be trained, ML has the potential to bypass the KS system altogether by explicitly modelling the Hohenberg-Kohn (HK) universal functional $F[n]$ \cite{Li2016} or equivalently the total energy $E[n]$ \cite{Ryczko2019,Bogojeski2020,Brockherde2017,Ray2025,delRio2023,Chen2025}. As diagonalising the KS Hamiltonian is the primary computational bottleneck of most DFT calculations, these data-driven approaches could facilitate the simulation of much larger systems than are currently tractable \cite{Chandrasekaran2019}.

Efforts to apply ML to this broad range of physical quantities are justified by the universality of models such as kernel ridge regression (KRR) and neural networks (NNs) \cite{Akashi2025}; given sufficient parameters, such models can approximate any mapping to arbitrary precision. However, the mappings from the electron density to some properties have attributes that make them harder to learn accurately, such as non-analyticity \cite{Moreno2020}. Even analytic density functionals can be \textit{ill-conditioned}, wherein small variations in the density lead to large changes in the functional \cite{Martinetto2024,Akashi2025}. A low-capacity ML model, e.g., one with too few parameters, will fail to capture this rapidly varying behaviour and underfit the data. Conversely, a model expressive enough to learn an ill-conditioned functional must be trained on a large, high-quality dataset to prevent overfitting, increasing computational demand.  As such, ill-conditioned mappings are inherently more challenging from an ML perspective \cite{Hanneke2023}. 

\begin{figure}[htbp]
\centering  

\tikzset{every picture/.style={line width=0.75pt}} 

\begin{tikzpicture}[x=0.75pt,y=0.75pt,yscale=-1,xscale=1]

\draw  (90,6831) -- (320,6831)(113,6660) -- (113,6850) (313,6826) -- (320,6831) -- (313,6836) (108,6667) -- (113,6660) -- (118,6667)  ;
\draw    (120,6800) .. controls (160,6770) and (150,6810) .. (190,6780) ;
\draw    (190,6780) -- (190,6720) ;
\draw    (190,6720) .. controls (267.2,6731.04) and (239.2,6703.04) .. (290,6700) ;
\draw  [dash pattern={on 0.84pt off 2.51pt}]  (190,6780) -- (113,6780) ;
\draw  [dash pattern={on 0.84pt off 2.51pt}]  (190,6720) -- (113,6720) ;
\draw  [dash pattern={on 0.84pt off 2.51pt}]  (190,6830) -- (190,6780) ;
\draw  (324,6831) -- (554,6831)(347,6660) -- (347,6850) (547,6826) -- (554,6831) -- (547,6836) (342,6667) -- (347,6660) -- (352,6667)  ;
\draw    (354,6800) .. controls (394,6770) and (370,6819) .. (410,6789) ;
\draw    (440,6722) .. controls (497.86,6715.33) and (473.2,6703.04) .. (524,6700) ;
\draw  [dash pattern={on 0.84pt off 2.51pt}]  (420,6830) -- (420,6775) ;
\draw  [dash pattern={on 0.84pt off 2.51pt}]  (427,6830) -- (427,6731) ;
\draw  [dash pattern={on 0.84pt off 2.51pt}]  (347,6775) -- (420,6775) ;
\draw  [dash pattern={on 0.84pt off 2.51pt}]  (347,6731) -- (424,6731) ;
\draw    (410,6789) .. controls (433.86,6770.33) and (408.86,6728.33) .. (440,6722) ;

\draw (121,6679) node [anchor=north west][inner sep=0.75pt]    {Density functional};
\draw (261,6799) node [anchor=north west][inner sep=0.75pt]    {Density};
\draw (93,6774) node [anchor=north west][inner sep=0.75pt]    {$Q_{1}$};
\draw (92,6715) node [anchor=north west][inner sep=0.75pt]    {$Q_{2}$};
\draw (185,6835) node [anchor=north west][inner sep=0.75pt]    {$\bar{n}=n$};
\draw (355,6675) node [anchor=north west][inner sep=0.75pt]    {Density functional};
\draw (499,6805) node [anchor=north west][inner sep=0.75pt]    {Density};
\draw (410,6835) node [anchor=north west][inner sep=0.75pt]    {$n_1$};
\draw (425,6835) node [anchor=north west][inner sep=0.75pt]    {$n_2$};
\draw (288,6669) node [anchor=north west][inner sep=0.75pt]   [align=left] {(a)};
\draw (529,6669) node [anchor=north west][inner sep=0.75pt]   [align=left] {(b)};
\draw (327,6770) node [anchor=north west][inner sep=0.75pt]    {$Q_{1}$};
\draw (327,6726) node [anchor=north west][inner sep=0.75pt]    {$Q_{2}$};

\end{tikzpicture}

\caption{(a) Schematic of a globally gauge-dependent density functional that experiences a constant shift in the external potential. The quantity jumps in value, but the electron density does not change. Hence, the change in the functional is discontinuous with respect to the density. As the same input maps to two different outputs, the learning problem is intractable. This situation is avoided by fixing the global gauge, which prevents constant shifts from occurring. (b) Schematic of a globally gauge-dependent density functional that experiences a \textit{near-constant} shift in the external potential. Although the functional is now smooth, machine-learning the mapping between the density and the functional is challenging in this region of density space because the functional changes so rapidly.}
\label{fig:constant_shift}
\end{figure}
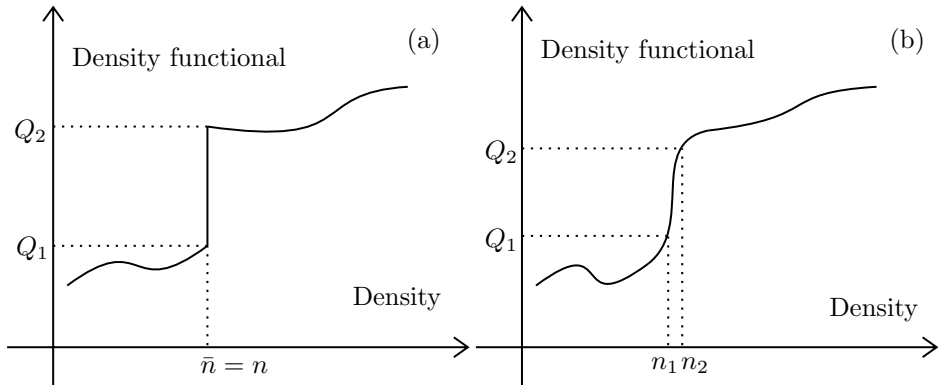

In this paper, we identify significant ill-conditioning in density functionals that belong to two classes, each of which contains properties of central interest in DFT. The first class corresponds to physical quantities that change value when a non-zero constant, $c$, is added to the external potential. We define these `globally gauge-dependent' quantities as follows:
\begin{definition}\label{def:global gauge-dependence}
Let the density $\bar{n}(\bm{r})$ and the physical quantity $\bar{Q}$ correspond to the potential $\bar{v}_\mathrm{ext}(\bm{r})$, and, likewise, $n(\bm{r})$ and $Q$ correspond to $v_\mathrm{ext}(\bm{r})$, where $\bar{v}_\mathrm{ext}(\bm{r}) = v_\mathrm{ext}(\bm{r}) + c$.

$Q$ is \textit{globally gauge-dependent} if $\bar{Q} \neq Q$, and \textit{globally gauge-invariant} if $\bar Q = Q$.
\end{definition}
Examples of globally gauge-dependent quantities include the total energy, \(E\), (\(\bar{E}=E+Nc\)), the first ionisation energy, $I$, (\(\bar{I}=I-c\)) and the electron affinity, $A$, (\(\bar{A}=A-c\)). The electron density itself, on the other hand, is globally gauge-invariant, i.e., $\bar n(\bm{r})=n(\bm{r})$. Therefore, a constant shift in the external potential will cause a globally gauge-dependent quantity to vary discontinuously with respect to the electron density, as depicted in Figure~\ref{fig:constant_shift}(a).
Due to the discontinuity, the mapping cannot be interpolated. However, this can be resolved by using a fixed global gauge, such as $v_\mathrm{ext}(r \rightarrow \infty) = 0$, which prevents any two potentials from differing exactly by a constant over all space. Under this constraint, the HK theorems \cite{Hohenberg1964} guarantee a one-to-one correspondence between $n(\bm{r})$ and $Q$, restoring the learnability of the functional.

However, we note that even in a fixed gauge, two potentials may differ approximately\footnote{The difference could even be exactly constant over a strict subset of space, while still satisfying the fixed gauge. However, this is unlikely for realistic systems.} by a constant in the region where the density is concentrated
(for example, see Fig.~\ref{fig:shift_example}). Since the electron density is globally gauge-invariant, \(n(\bm{r})\) and \(\bar n(\bm{r})\) are likely to be very similar if the difference between $v_\mathrm{ext}(\bm{r})$ and $\bar v_\mathrm{ext}(\bm{r})$ is near-constant. On the other hand, the globally gauge-dependent quantities \(Q\) and \(\bar Q\) may differ significantly, e.g., \(\bar E - E \approx Nc\). $Q[n]$ is therefore ill-conditioned, as similar inputs correspond to highly disparate outputs (this is illustrated in Figure~\ref{fig:constant_shift}(b)). In summary, an exactly constant shift causes globally gauge-dependent quantities to vary discontinuously with respect to the electron density, making learning impossible; this is avoided by fixing the global gauge. However, a near-constant shift, which can occur even in a fixed global gauge, causes globally gauge-dependent quantities to vary rapidly with respect to the electron density, making learning highly challenging.

The second class of ill-conditioned functionals that we highlight are functionals of the $N$-electron density that depend strongly on the properties of the corresponding $(N+1)$-electron system. Examples include the electron affinity, $A$, and the fundamental gap, $E_g$, which both explicitly depend on the total energy of the $(N+1)$-electron system ($A=E_N-E_{N+1}$ and $E_g=E_{N-1}-2E_N+E_{N+1}$). Ill-conditioning arises if some change to the external potential significantly affects the $(N+1)$-electron system, and hence the functional output, but not the $N$-electron system, and hence not the functional input. Consider two external potentials that only differ significantly in one region of space; for example, see Figure~\ref{fig:second_example}(c). If the $N$-electron densities of both systems are low in this region, they are likely to be insensitive to the difference and hence similar overall. However, if the corresponding $(N+1)$-electron systems have significant density in the region where the external potentials differ appreciably, properties such as the affinity and fundamental gap will differ significantly for the two external potentials. These quantities are therefore ill-conditioned when expressed as functionals of the $N$-electron density.

Below, we use one-dimensional (1D) model systems to demonstrate the challenge of applying ML to these two classes of density functionals. We use the iDEA code \cite{PhysRevB.88.241102} to generate a dataset comprised of various systems of two fully interacting electrons on a real-space ($x$) grid by solving the many-body Schr\"odinger equation numerically. For each system, we calculate the exact many-body electron density $n(x)$, the total energy $E$ (Class 1), the first ionisation energy $I$ (Class 1), the electron affinity $A$ (Classes 1 and 2), the fundamental gap $E_g$ (Class 2), and the HK universal functional $F$. Using finite difference quotients, we demonstrate that quantities in the two classes vary extremely rapidly with respect to the electron density. We contrast this with the universal functional, which falls in neither class and subsequently does not exhibit ill-conditioning in our dataset.

We then train KRR models to predict each physical quantity, providing the entire (discretised) electron density as the input, and observe several orders of magnitude greater error for all quantities compared to $F$. We also identify global gauge-dependence as an explanation for ML error in previous work \cite{Ryczko2019}, indicating a problem that persists across a variety of physical systems and ML techniques.

In addition to highlighting these sources of ill-conditioning as potential pitfalls for accurate ML in DFT, we also propose practical solutions. The ill-conditioning of the total energy functional $E[n]$ is easily circumvented by instead machine-learning the universal functional $F[n]$ and adding the external potential energy component. We also demonstrate that using the external potential as input to the ML model can improve accuracy for \textit{all} quantities in the two classes. This solution is based on the observation that the small changes in electron density characteristic of ill-conditioning correspond to significant changes in the external potential, so the quantities are not ill-conditioned when expressed as potential functionals. We validate this approach by training KRR models to predict all of the aforementioned quantities using the external potential as input instead of the electron density, and observe significantly greater accuracy for properties in both Classes 1 and 2.

\section{Methods}\label{sec2}
\subsection{Data generation}

To obtain our training and testing data, we start by defining one hundred 1D molecular potentials of the form 
\begin{equation}
    v_\mathrm{ext}(x)=-\frac{Z_l}{\sqrt{\left (x+\frac{R}{2} \right )^2+\alpha_l^2 }}-\frac{Z_r}{\sqrt{\left (x-\frac{R}{2} \right )^2+\alpha_r^2}}.
    \label{eq:vatom}
\end{equation}
(Atomic units are used throughout.) These potentials have the soft-core Coulomb form, which uses a softening parameter, \(\alpha\) (since the 1D bare-Coulomb form has a non-integrable singularity) \cite{Gordon2005,Loudon2016}. The global gauge is fixed, as $v_\mathrm{ext}(|x| \rightarrow \infty)=0$ for all potentials. We use a fixed internuclear separation of $R=5$ and sample the remaining parameters uniformly in the following ranges: $3 \leq Z_l \leq3.5$, $1.25 \leq \frac{1}{\alpha_l} \leq1.75$, $1.5 \leq Z_r \leq2$, $1.25 \leq \frac{1}{\alpha_r} \leq1.75$. The external potentials are represented on a spatial grid of 75 points from $x=\SI{-7.5}{a_0}$ to $x=\SI{7.5}{a_0}$, ensuring that all relevant physical quantities are converged to within \SI{e-5}{Ha}.
For each external potential, we model systems of two electrons in a spin singlet (spin up and spin down), interacting via the soft-core Coulomb potential with \(\alpha=1\), i.e., $U(x,x')=\left ((x-x')^2+1 \right )^{-\frac{1}{2}}$. We numerically solve the many-body Schr\"odinger equation using iDEA to find the exact total energy, \(E\), and the many-body wavefunction, from which we obtain the exact many-body density, \(n(x)\). We calculate the value of the HK universal functional for each system by subtracting the external potential energy from the total energy: \(F=E-E_v\), where $E_v=\int_{-\infty}^{\infty}n(x)v_\mathrm{ext}(x)\mathrm{d}x$. 
To obtain the remaining quantities, we first solve the Schr\"odinger equation for the corresponding one- and three-electron systems\footnote{For the three-electron systems, our electron spins are up, up, and down.}. We then calculate the first ionisation energy, \(I=E_{1}-E_{2}\), the electron affinity, \(A=E_2-E_3\), and the fundamental gap, \(E_\mathrm{g}=I-A\). 

\subsection{Data analysis}\label{sec22}

\begin{figure}[h]
    \centering
    \includegraphics[width=1\linewidth]{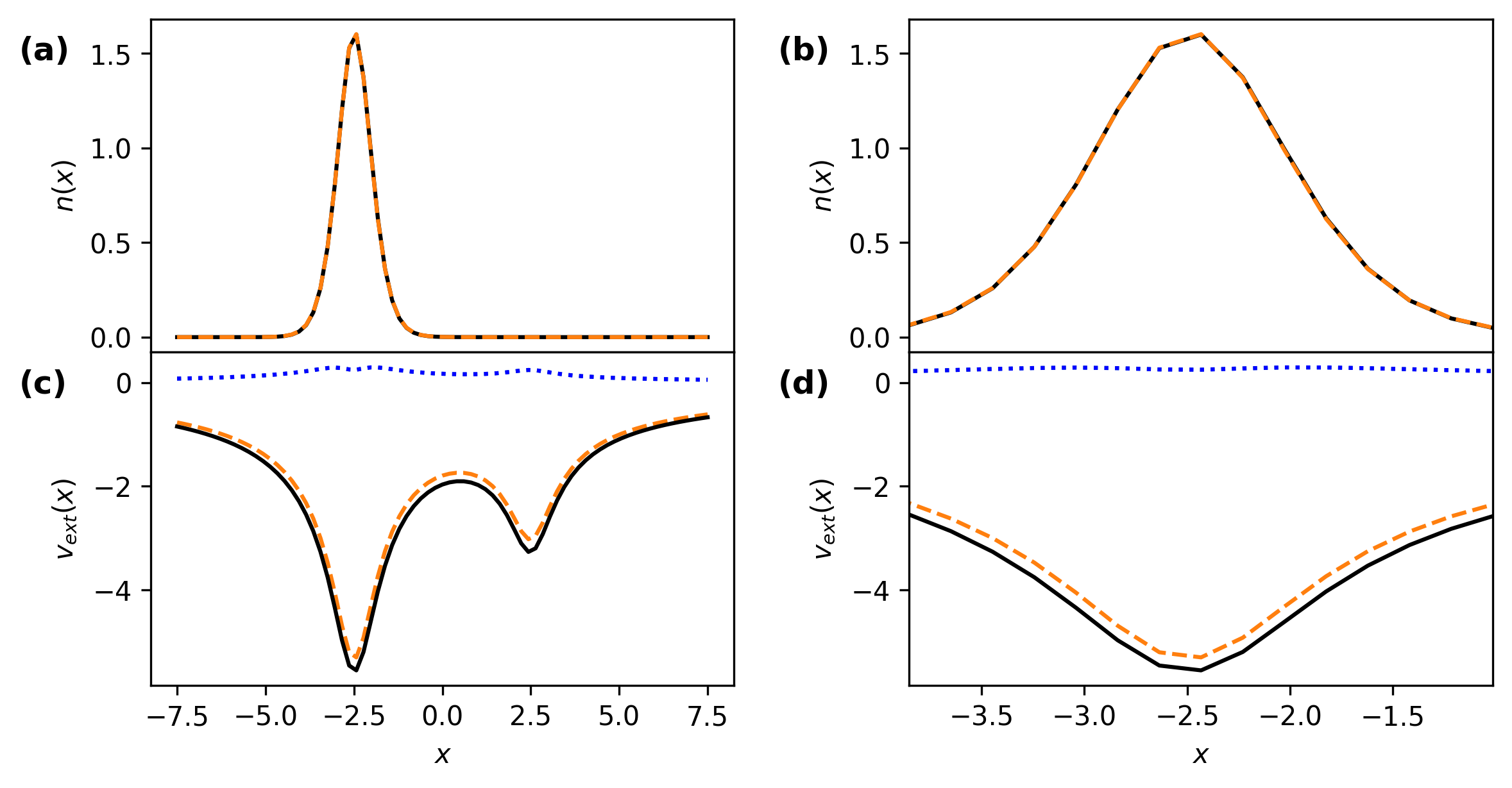}
    \caption{An example pair of two-electron molecular systems from our dataset that shows how systems can differ by a near-constant shift. The black solid lines and dashed orange lines each correspond to a system (see text for details), for which the external potentials and electron densities are plotted in (c) and (a), respectively. The blue dotted lines show the difference between the external potentials, which, due to the fixed global gauge, is non-constant. However, panel (d) shows that over the region where most of the electron density is concentrated, the difference between the two external potentials is approximately constant. Consequently, the two systems have almost identical electron densities (b), but differ substantially in their globally gauge-dependent properties -- for example, the system represented by solid black lines has a total energy of -7.72 Ha, whereas the other system has a total energy of -7.18 Ha.}
    \label{fig:shift_example}
\end{figure}

We now illustrate the ill-conditioning effects described above using examples from our dataset. Figure~\ref{fig:shift_example} shows a pair of systems that exemplifies the ill-conditioning of globally gauge-dependent quantities. The molecular potentials in panel (c) are defined by \(Z_l=3.44,\alpha_l=0.65,Z_r=1.64,\alpha_r=0.63\) (solid black line) and \(Z_l=3.10,\alpha_l=0.62,Z_r=1.52,\alpha_r=0.63\)  (dashed orange line). Their difference, $\bar v_\mathrm{ext}(x)-v_\mathrm{ext}(x)$, is also shown as a blue dotted line. $\bar v_\mathrm{ext}(x)-v_\mathrm{ext}(x)$ is non-constant, peaking at $x=\pm R/2$ and tending to 0 as \(|x| \rightarrow \infty\) (as necessitated by the fixed global gauge). However, over the domain $x=-3.9$ to $x=-1$ (right), which contains over 97\% of the electron density, it is approximately constant. As a result, the corresponding electron densities, \(n(x)\) and \(\bar n(x) \), are very similar. This is visually apparent in panel (b), but we also attain a numerical measure using the \(L^2\) norm distance,
\begin{equation*}\label{eq:L2}
    ||\bar n(x)-n(x)||_{L^2}=\sqrt{\int_{-\infty}^{\infty}\left [ \bar n(x)-n(x) \right ]^2 \mathrm{d} x},
\end{equation*}
and find that \(||\bar n(x)-n(x)||_{L^2}=0.004\). Despite their proximity in density space, the difference in total energy is significant at \(\bar E-E=0.54\); the total energy functional, \(E[n]\), is ill-conditioned.
In contrast, the value of the HK universal functional \(F\) differs by only 0.004 for the two systems, so $F[n]$ is not ill-conditioned. Since $E=F+E_v$, this shows that, as expected, the gauge-dependent energy component $E_v$ is responsible for the significant difference in total energy. The near-constant shift also leads to large changes in $I$ (0.27) and $A$ (0.21), because they are globally gauge-dependent.

The rate at which a functional \(Q[n]\) changes with respect to the input can be approximated by the finite difference quotient
\begin{equation*}
\frac{\Delta Q}{\Delta n}=\frac{|Q[\bar n]-Q[n]|}{||\bar n(x)-n(x)||_{L^2}}.
\end{equation*}
$\Delta Q/\Delta n \gg1$ indicates ill-conditioning, as $\Delta n$ is small relative to $\Delta Q$. For the systems shown in Figure~\ref{fig:shift_example}, \(\Delta E/\Delta n=122\) and \(\Delta F/\Delta n=0.96\). 

\begin{figure}[h]
    \centering
    \includegraphics[width=0.6\linewidth]{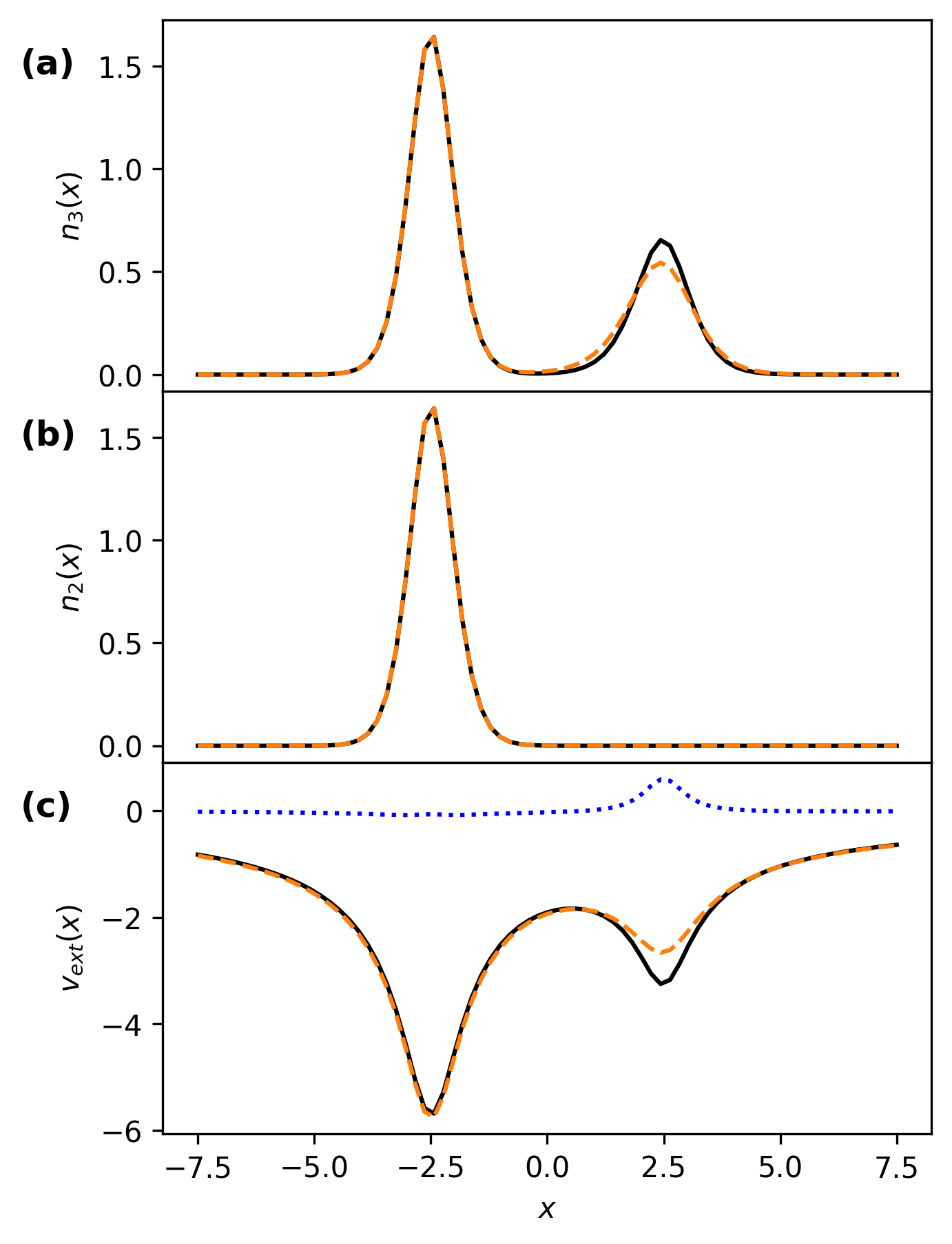}
    \caption{Two systems (one indicated by solid black lines, the other by dashed orange lines -- details in the main text) that illustrate how an implicit dependence on the \(\left ( N+1 \right )\)-electron density can lead to ill-conditioning. They are both two-electron molecular systems wherein both electrons are confined to the left well (b). In this region, the difference between the external potentials of the two systems (c) is close to 0, leading to similar electron densities. However, the corresponding three-electron densities (a) are spread between both wells, and the potentials differ significantly in the right well. This leads to large differences between the three-electron densities and quantities that depend on them, such as the affinity and fundamental gap. As a consequence, the changes in these properties of the two-electron systems are disproportionate to the small change in the two-electron density.}
    \label{fig:second_example}
\end{figure}

Figure~\ref{fig:second_example} shows a pair of systems from our dataset that exemplifies the ill-conditioning of our second class of functionals, i.e., any functional of the $N$-electron density with a strong implicit dependence on the $(N+1)$-electron density, such as $A[n]$ and $E_g[n]$. The potentials (c) are defined by \(Z_l=3.49,\alpha_l=0.64,Z_r=1.52,\alpha_r=0.77\) (solid black line) and \(Z_l=3.39,\alpha_l=0.63,Z_r=1.52,\alpha_r=0.59\)  (dashed orange line). The two-electron densities (b) are concentrated in the left wells, where the difference between the potentials (blue dotted line) is close to 0.  As a result, they essentially experience the same potential and differ by only 0.001 in $L^2$.
However, the difference between the potentials is significant in the right well, which is within the domain of the three-electron densities (a). As such, $E_g$ and $A$ for the two systems differ by 0.39 and 0.32 respectively, with finite difference quotients of $\Delta E_g/\Delta n =318$ and $\Delta A/\Delta n =259$. In contrast, $\Delta F/\Delta n=0.83$.

We now extend this analysis to all pairs of systems in our dataset $i$ and $j$, $i \ne j$. For each pair, we compute the finite difference quotient \(\Delta Q / \Delta n\) for all quantities. We then take the maximum finite difference quotient for each quantity as a rough empirical estimate of the local Lipschitz constant, \(\hat L\) \cite{Huang2023on}.

\begin{equation}\label{eq:Lipschitz}
\hat{L}=\text{max}_{i \ne j} \frac{|Q[n_i]-Q[n_j]|}{||n_i(x)-n_j(x)||_{L^2}}  
\end{equation}

$\hat L$ is a lower bound on the maximum rate of change within the domain of the dataset. It is commonly used as a measure of learning complexity \cite{Ashlagi2024}, e.g., to attain bounds for the worst-case prediction error \cite{Huang2023on}. We use it to characterise the degree of ill-conditioning of each functional in our dataset.

\begin{figure}[h]
    \centering
    \includegraphics[width=0.75\linewidth]{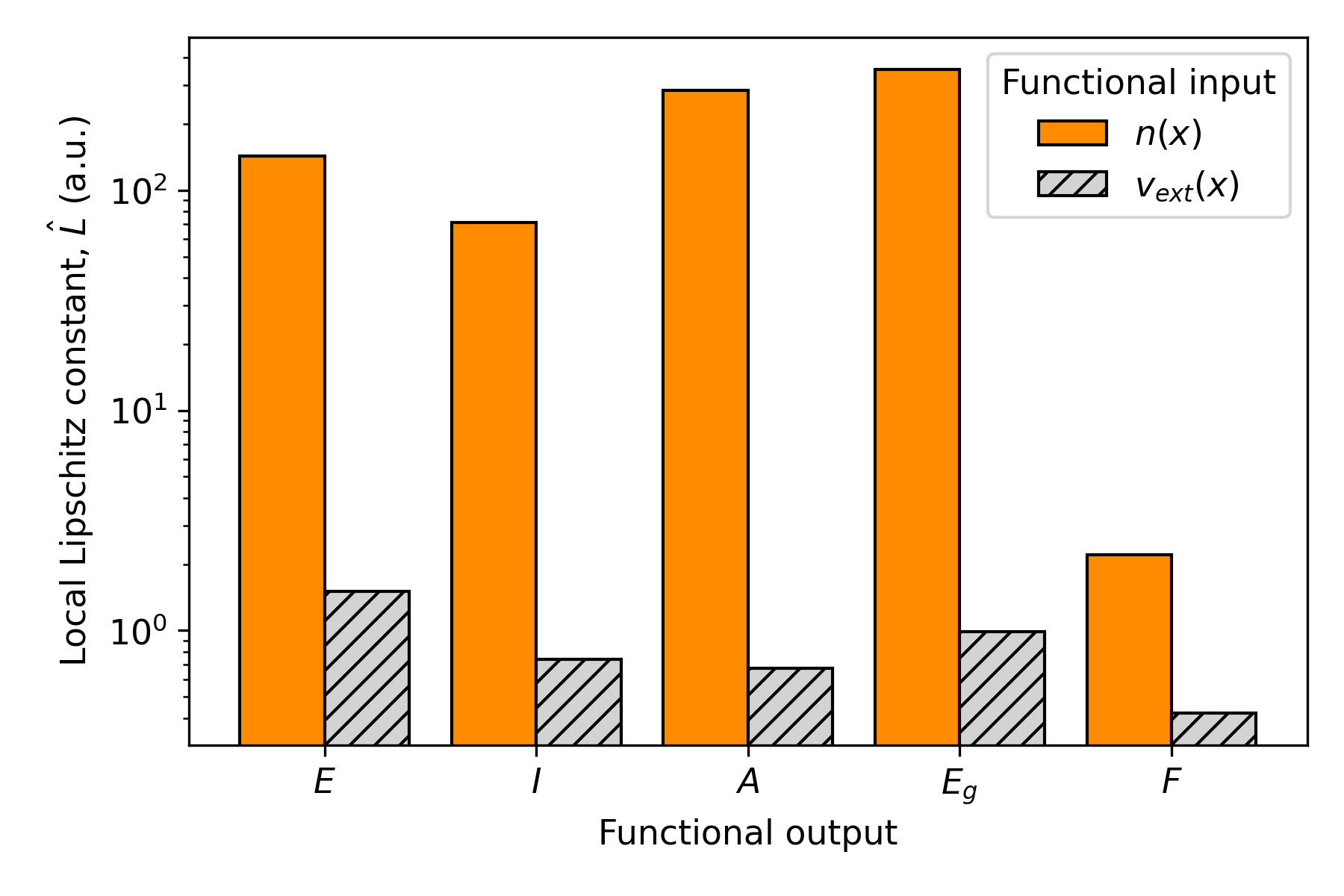}
    \caption{The local Lipschitz constant \(\hat L\) is a lower bound on the maximum of a function or functional's rate of change within a restricted domain. We attain a simple estimate by computing the maximum finite difference quotient between all pairs of systems in our dataset. \(\hat L\) is calculated for all density functionals (orange bars) and potential functionals (grey hatched bars), enabling a comparison of how rapidly each varies with respect to its input. The ill-conditioning effects we have identified manifest as values of \(\hat L\) on the order of \(10^2\) for \(E[n]\), \(I[n]\), \(A[n]\) and \(E_g[n]\). For \(F[n]\), which does not suffer from these effects, \(\hat L\) is two orders of magnitude smaller. Similarly, \(\hat L\) is on the order of \(10^0\) for all potential functionals, indicating the absence of ill-conditioning. }
    \label{fig:Lipschitz}
\end{figure}

In Figure~\ref{fig:Lipschitz} we plot \(\hat L\) for all density functionals (solid orange bars) in our dataset. $\hat L\gg 1$ for $E[n]$, $I[n]$, $A[n]$ and $E_g[n]$, providing numerical confirmation of the ill-conditioning described above. We also plot $\hat L$ for each quantity as a functional of the external potential (grey hatched bars), showing that the ill-conditioning does not arise for potential functionals. This is because the small changes in electron density in general correspond to significant changes in the external potential (for example, see Figures~\ref{fig:shift_example} and \ref{fig:second_example}), so $\Delta Q/\Delta v_\mathrm{ext}$ is consistently small. 

\subsection{Machine learning}

KRR is a non-linear regression model with regularisation \cite{Li2014}. To approximate a functional $Q[n]$, KRR uses a kernel function \({k[n_i,n_j]}\) that measures the similarity of two inputs \(n_i\) and \(n_j\). We use the Gaussian kernel, $k[n_i,n_j]=\exp(-||n_i-n_j||_{L^2}^2/(2\sigma^2))$, a standard choice in ML-DFT \cite{Snyder2012,Brockherde2017,Li2014,Vu2015}. For a given input, $n$, the predicted value, $Q^\mathrm{ML}[n]$, is calculated as a weighted sum of the similarity scores of $n$ with each $n_j$ in the training set \cite{Vu2015}, of which there are $M$, i.e., 
\begin{equation*}\label{eq:KRR_form}
    Q^\mathrm{ML}[n]=\sum_{j=1}^{M} \alpha_jk[n,n_j] . 
\end{equation*}
The weights \(\bm {\alpha}=(\alpha_1,...,\alpha_{M})^T\) are found by minimising the cost function
\begin{equation*}\label{eq:KRR_cost_function}
    C(\bm\alpha)=\sum_{j=1}^{M} (Q^\mathrm{ML}[n_j]-Q[n_j])^2 +\lambda ||Q||_{\mathcal{H}_c}^2, 
\end{equation*} 
where \(||Q||_{\mathcal{H}_c}^2=\bm{\alpha} \cdot\mathbb{K} \bm{\alpha} \) is the reproducing kernel Hilbert space (RKHS) norm \cite{Kanagawa2018} and \(\mathbb{K}\) is a matrix with elements \(K_{ij}=k[n_i,n_j]\). The greater the value of the regularisation parameter $\lambda$, the more emphasis is placed on minimising the RKHS norm, which can prevent the model from overfitting the training data.

The cost function is minimised by solving $\bm {\alpha}=(\mathbb{K}+\lambda \mathbb{I})^{-1} \bm{Q}$, where \(\mathbb{I}\) is the identity matrix and \(\bm{Q}=(Q_1,...,Q_{M})^T\) are the outputs (or `labels') in the training set \cite{Vu2015}. Note that both \(\lambda\) and \(\sigma\) are hyperparameters, meaning their values are chosen explicitly before training. The small number of hyperparameters in KRR makes optimising (or `tuning') them simpler than in more complex models such as NNs. Training is also easier for KRR due to the closed-form solution, which finds the global minimum of the loss function. In contrast, there is no such guarantee for NNs, which are typically trained via stochastic optimisation methods. As such, we employ KRR in this paper as the technique relies on user input less than NNs, offering a fairer demonstration of our hypotheses. However, we note that our findings are not specific to KRR, as we observe similar results in previous work using NNs; see Ref.~\cite{Ryczko2019} (discussed below). 

To tune and test our KRR models, we use nested k-fold cross-validation in combination with grid searches \cite{Hansen2013}. The nested validation comprises 10 outer folds and 9 inner folds with an 80-10-10 train-validation-test split, providing robust measures of model performance both at the validation and testing stages. Previous studies have shown that KRR can learn highly accurate DFAs from a small number of training samples \cite{Snyder2013}, and may outperform NNs in the small-data regime \cite{Hansen2013}.

To assess model accuracy, we use the mean absolute error (MAE)
\begin{equation*}\label{eq:MAE}
   \mathrm{MAE}=\frac{1}{S}\sum_{i=1}^{S}|Q^\mathrm{ML}[n_i]-Q[n_i]|,
\end{equation*}
where $n_i$ are the inputs in the test set, of which there are $S=10$ (note that in the nested k-fold cross-validation, all 100 samples are used for testing at some point).

\section{Results}\label{sec3}

\begin{figure}[h]
    \centering
    \includegraphics[width=0.75\linewidth]{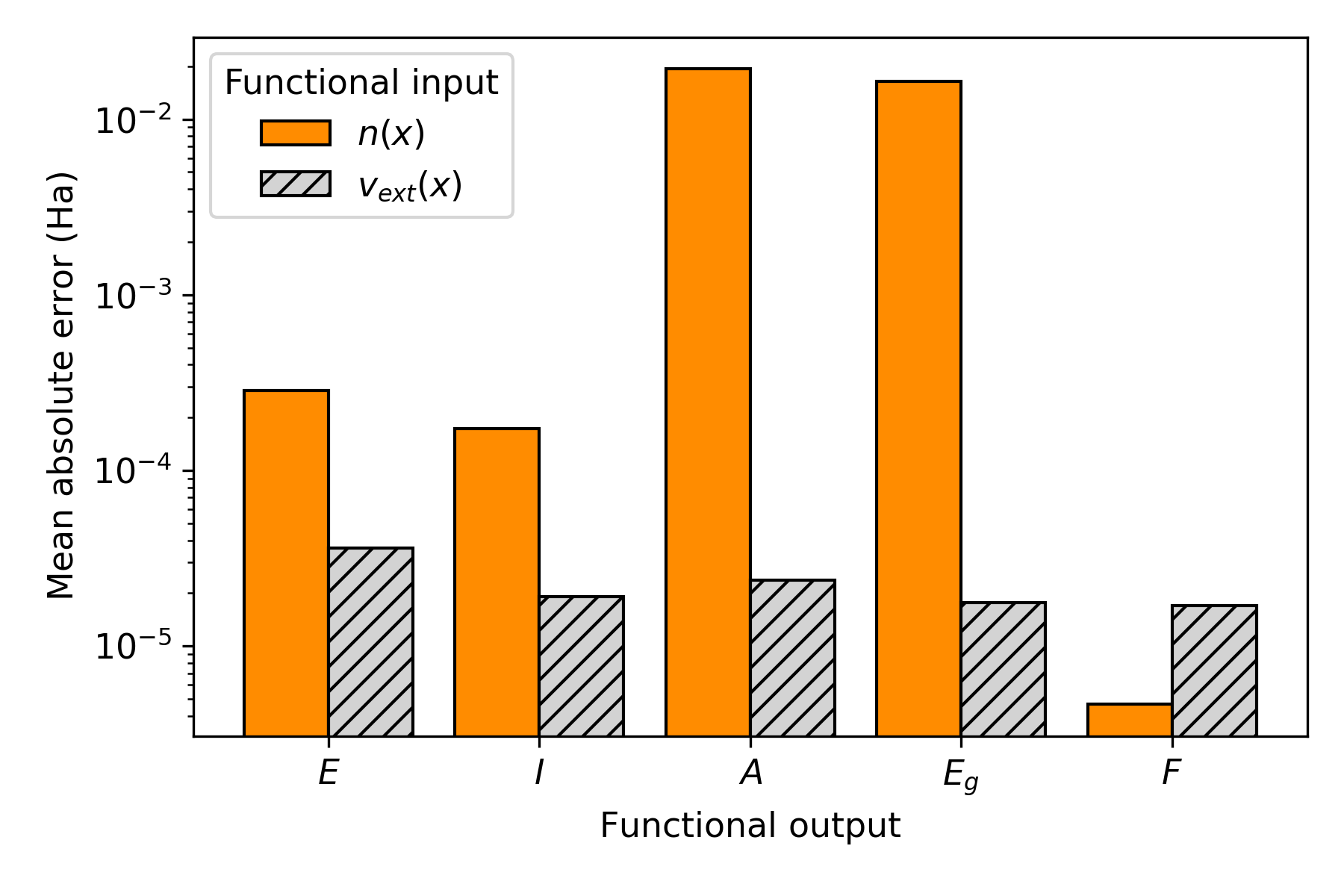}
    \caption{The MAE in our ML approximations to density functionals (orange bars) and potential functionals (grey hatched bars). For the density functionals, the error spans several orders of magnitude: from \(10^{-6}\) Ha ($F[n]$) to \(10^{-4}\) Ha ($E[n]$ and $I[n]$) to \(10^{-2}\) Ha ($A[n]$ and $E_g[n]$). In contrast, the error is far more balanced between the different potential functionals, on the order of \(10^{-5}\) Ha for all quantities. This is in close parallel to the pattern in Figure~\ref{fig:Lipschitz}, where we observe much greater Lipschitz constants for other density functionals compared to $F[n]$, but more of a balance in the potential functionals. As such, we can infer that ill-conditioning is a primary determinant of the error in these results, explaining the significantly lower accuracy observed for $E[n]$, $I[n]$, $A[n]$, and $E_g[n]$.}
    \label{fig:mae}
\end{figure}

Figure~\ref{fig:mae} shows the MAE in the KRR predictions of each physical quantity when using the electron density (solid orange bars) and the external potential (hatched grey bars) as input. Amongst the density functionals, the error is orders of magnitude greater for the functionals we identified as ill-conditioned ($E[n]$, $I[n]$, $A[n]$, and $E_g[n]$). For example, although the error in $E[n]$ was very low in absolute terms (0.28 mHa or 0.18 kcal/mol) (due to the quality of data provided by our 1D systems), it was 61$\times$ greater than the error in $F[n]$ ($4.65 \times10^{-3}$ mHa or $2.9\times10^{-3}$ kcal/mol). The error in $A[n]$ and $E_g[n]$ was the greatest by a significant margin at 19.4 and 16.4 mHa, respectively -- around 4000$\times$ greater than the error in $F[n]$. This is likely because these quantities, due to their dependence on the $(N+1)$-electron density, exhibit the most extreme ill-conditioning in the dataset (note that they have the greatest values of $\hat L$ in Figure~\ref{fig:Lipschitz}).
 
The total energy, first ionisation energy, electron affinity, and fundamental gap are all predicted with significantly greater accuracy from the external potential, because the ill-conditioning issue is avoided. Specifically, we observe about a 10-fold decrease in error for $E$ and $I$, and a 1000-fold decrease for $A$ and $E_g$, when using the external potential as input rather than the electron density. This confirms that the pathologies of the density\(\rightarrow\)output mapping are the source of error, not the output quantities \textit{per se.} 

Our results are in striking agreement with Reference~\cite{Ryczko2019}, where ML was employed to predict the total energy, external potential energy, exchange, correlation and kinetic energy of electrons in random 2D potentials. 
When using the electron density as input to their ML models, the authors observed significantly greater error for the total energy and external potential energy relative to the exchange, correlation, and kinetic energy; the MAE was typically an order of magnitude greater. We posit that this is explained by ill-conditioning, as $E$ and $E_v$ are globally gauge-dependent and the other quantities are not. However, the authors attributed it to the fact that their dataset contained a larger range of values for $E$ and $E_v$ compared to the other quantities. As this explanation relies solely on the output property, it implies that the discrepancy in predictive accuracy should not depend on the input to the ML model. However, when the authors used the external potential as the input, the error was more balanced between the different properties. Likewise, we find that when using the external potential as the input, our KRR model is equally capable of predicting globally gauge-dependent quantities and globally gauge-invariant quantities; see Figure~\ref{fig:mae}. Our ill-conditioning hypothesis explains why the discrepancy occurs for density functionals and not potential functionals, both in our results and Reference~\cite{Ryczko2019}. In addition, as the authors of Reference~\cite{Ryczko2019} used NNs, while we used KRR, this implies the difficulties of learning these density functionals are independent of the ML method. The sensitivity of NNs to this ill-conditioning could be explained by the `spectral bias' of NNs, which prioritise learning low-frequency components \cite{Rahaman2019}.

As we have highlighted, these ill-conditioning effects stem from specific changes in the external potential, such as near-constant shifts. If these regions of functional space can be avoided, then so can the associated machine learning challenges. However, the central aim of ML-DFT is to train a model that can reliably generalise across diverse physical systems \cite{Akashi2025}; this requires a broad and varied dataset, so avoiding ill-conditioning may not be feasible in practice. For example, Reference~\cite{Ryczko2019} used a very broad dataset, so it is not surprising that these effects were encountered. This is why an alternative solution, such as using the external potential as input, is required. 

\section{Conclusion}\label{sec4}

Although in principle, ML models such as kernel ridge regression (KRR) and neural networks (NNs) can learn the mapping from the electron density to any electronic property, some physical quantities can be highly sensitive to small variations in the electron density, which makes accurate machine learning far more challenging. In this paper, we identified two sources of this `ill-conditioning' within density functional theory (DFT), each affecting 
an entire class of physical quantities: 1) Globally gauge-dependent quantities (defined as quantities which vary when a constant shift is applied to the external potential), and 2) functionals of the $N$-electron density that have a strong implicit dependence on the $(N+1)$-electron density.  

These ill-conditioning effects are caused by specific changes to the external potential of a system that have only a negligible impact on its electron density but a disproportionately large impact on the target quantity. This occurs in the first class of functionals when the potential shift is near-constant (which is possible even when the global gauge is fixed), causing quantities such as the total energy to vary rapidly while the electron density is largely unaffected. Ill-conditioning arises in the second class of quantities when the change in potential is small over the domain of the $N$-electron density but significant in the domain of the $(N+1)$-electron density. The density of the $N$-electron system responds minimally, whereas any of the system's properties that have a strong dependence on the $(N+1)$-electron density, such as the electron affinity, are significantly impacted.

Using exactly solvable one-dimensional model systems that form the basis of our ML dataset, we provide examples of these effects, where pairs of systems with similar electron densities differ substantially in their total energy (Class 1), first ionisation energy (Class 1), electron affinity (Classes 1 and 2) or fundamental gap (Class 2).
We find that with a fixed dataset size and training procedure, KRR models exhibit significantly greater error when applied to these functionals compared to the Hohenberg-Kohn universal functional, which does not suffer from the sources of ill-conditioning we identified. 
Both in this research and previous work which employed NNs \cite{Ryczko2019}, the external potential energy, $E_v$, has proven to be the most challenging component of $E[n]$ to model with ML. Our analysis explains this surprising result; although $E_v[n]$ is physically simple, it is ill-conditioned and hence highly complex from an ML standpoint. Since \(E_v\) can be calculated via a closed-form expression, our findings highlight that it is preferable to machine-learn \(F[n]\) and then add the external potential energy contribution to obtain a reliable approximation to the total energy.

In addition, we find that none of the ill-conditioning identified in this paper is present in the potential functionals corresponding to these quantities. As a consequence, circumventing the electron density altogether and using the external potential as the input to the ML model led to far more accurate results. 

Although our numerical examples are one-dimensional, our arguments readily extend to two- and three-dimensional systems.
Moreover, while we have focused on two specific sources of ill-conditioning, our solution of using the external potential as input to the ML model would also circumvent other possible sources of ill-conditioning in density functionals that we have not identified in this paper. Future research could apply our methodology to identify ill-conditioning in realistic ML-DFT datasets and to determine whether the potential-based approach could be beneficial. As such, our findings offer the prospect of reliable ML models within DFT for a variety of important physical quantities that may, in turn, lead to more advanced materials simulations.

\section*{Declarations}



\bmhead{Acknowledgements}
This study was funded by the N8 Centre of Excellence in Computationally Intensive Research (funded by UKRI, grant reference EP/T022167/1). The funder played no role in study design, data collection, analysis and interpretation of data, or the writing of this manuscript. P.J.H. acknowledges funding from UKRI (grant references UKRI2710, EP/W030489/1 and EP/X035891/1).

\bmhead{Data availability}
The datasets generated and analysed during the current study are openly available in the research data repository of the University of York, https://doi.org/10.15124/be174a35-5ae0-4758-85f1-d21d7cffdeba. 




\bmhead{Code Availability}

The underlying code for this study is available in the iDEA-org GitHub repository and can be accessed via this link \href{https://github.com/iDEA-org/iDEA}{https://github.com/iDEA-org/iDEA}.

\bmhead{Author contribution}

L.A. performed all numerical calculations, including the development of machine learning (ML) software and the training of all ML models. Under the supervision of M.J.P.H. and R.L., L.A. identified the sources of numerical ill-conditioning. M.J.P.H. and J.W. conceptualised the project, with M.J.P.H. directing the research and proposing the focus on constant shifts in the potential. M.J.P.H. and J.W. authored the iDEA software, implementing substantial improvements that directly enabled this work; J.W. further extended iDEA with GPU acceleration for many-body interacting systems. R.L. and P.J.H. provided insights into DFT and ML, proposed conceptual ideas, and contributed to discussions throughout the project. L.A. and M.J.P.H. drafted the manuscript, while all authors contributed to the review and editing of the manuscript and approved the final version.

\bmhead{Competing interests}
All authors declare no financial or non-financial competing interests.





\begin{appendices}




\end{appendices}

\bibliography{sn-bibliography}

\end{document}